\documentclass[preprint, pteplogo]{ptephy_v1} 

\preprintnumber{-}

\usepackage{amsmath}  
\usepackage{amsfonts} %
\usepackage{graphicx} 
\usepackage{url}

\begin{document}

\def\be{\begin{equation}}
\def\ee{\end{equation}\noindent}
\def\bear{\begin{eqnarray}}
\def\ear{\end{eqnarray}\noindent}
\def\bec{\blue\begin{equation}}
\def\eec{\end{equation}\black\noindent}
\def\bearc{\blue\begin{eqnarray}}
\def\earc{\end{eqnarray}\black\noindent}
\def\benn{\begin{enumerate}}
\def\enn{\end{enumerate}}
\def\veject{\vfill\eject}
\def\ven{\vfill\eject\noindent}

\title{The Yukawa potential: ground state energy and critical screening}

\author[1]{James P. Edwards}
\affil{Instituto de F\'isica y Matem\'aticas, Universidad Michoacana de San Nicol\'as de Hidalgo,
Edificio C$-3$, Apdo. Postal $2-82$, C.P. $58040$, Morelia, Michoac\'an, Mexico \email{schubert@ifm.umich.mx}}
\author[1,2]{Urs Gerber}
\affil{Instituto de Ciencias Nucleares, Universidad Nacional Aut\'onoma de M\'exico, A.P. 70-543, C. P. 04510, Ciudad de M\'exico, Mexico}

\author[1]{Christian Schubert}
\author[1]{Maria Anabel Trejo}
\author[1]{Axel Weber\thanks{All authors contributed equally to this work}}

\date{\today}

\begin{abstract}
We study the ground state energy and the critical screening parameter of the Yukawa potential in non-relativistic quantum mechanics. 
After a short review of the existing literature on these quantities, we apply fifth-order perturbation theory to the calculation of 
the ground state energy, using the exact solutions of the Coulomb potential together with a cutoff on the principal number summations.
We also perform a variational calculation of the ground state energy using a Coulomb-like radial wave function 
and the exact solution of the corresponding minimization condition. For not too large values of the screening parameter, close agreement is found between
the perturbative and variational results. For the critical screening parameter, 
we devise a novel method that permits us to determine it to ten digits.
This is the most precise calculation of this quantity to date, and allows us to resolve some discrepancies between previous results. 
\end{abstract}

\maketitle 

\section{Introduction}
\label{sec:intro}
\renewcommand{\theequation}{1.\arabic{equation}}
\setcounter{equation}{0}

The Yukawa potential was proposed by Yukawa in 1935 \cite{Yukawa} as an effective non-relativistic potential describing the
strong interactions between nucleons. It takes the form 

\begin{equation}
 \label{ec:yukawa}
 V(r) = -\alpha\frac{\textrm{e}^{-\mu r}}{r},
\end{equation}
and thus can be seen as a screened version of the Coulomb potential, with $\alpha$ describing the strength of the interaction and $1/\mu$ its range.
The same potential appears under the name of Debye-H\"uckel potential in plasma physics, where it represents the potential of a charged particle in a 
weakly nonideal plasma \cite{Debye}, as well as in electrolytes and colloids.
In solid state physics it is known as Thomas-Fermi potential, and  describes the effects of a charged particle in a sea of conduction electrons.

In quantum mechanics, the physics of this potential depend strongly on the value of the screening parameter $\mu$. While for the Coulomb case $\mu=0$ 
there is an infinite number of bound states, for any positive value of $\mu$ the screening is sufficient to reduce this number to a finite one
\cite{JostPais, Bargmann, Schwinger}, and for $\mu$ larger than a certain critical value $\mu_c$,  bound states cease to exist 
altogether. This critical value is proportional to \cite{JostPais, Harris, Rogers, Garavelli, Gomes, Leo, Yongyao} $\alpha m$:

\bear
\mu_c \approx 1.19 \, \alpha m\,.
\label{mucrit}
\ear
Despite its superficial closeness to the Coulomb potential, the Yukawa one shares hardly any of the exceptional mathematical properties of the former. 
To this date, for $\mu \ne 0$ neither the energy eigenvalues, nor the eigenfunctions, nor the critical screening parameter are known in closed form. 
This combination of physical importance and mathematical intractability makes the Yukawa potential a natural test case for approximation methods
in quantum mechanics. 

The purpose of the present paper is fourfold. First, in section \ref{sec:review} we will give an overview 
of the various approximation methods that have been used to date. Our emphasis here is on the ground
state energy $E_0(\mu)$ and on the critical screening $\mu_c$, since these are the quantities which we are then studying ourselves in the rest of the paper. 
We will use these literature data to construct a literature average curve $E_0(\mu)$.   

Second, in section \ref{sec:pt} we will perform a perturbative calculation of the ground state energy, taking the exactly solvable Coulomb Hamiltonian as the
unperturbed one. The special
properties of the Coulomb case will allow us to push this calculation to an unusual fifth order. 
We further improve on this calculation by including in the unperturbed Hamiltonian the
contribution from the Yukawa potential that is linear in $\mu$. We also study the dependence of the perturbative calculation for the energy on the cutoff 
in the principal quantum number of the Coulomb wave functions that becomes necessary starting 
from at second order. 

Third, in section 4 we then compare our perturbative results with a variational calculation, using a trial function of the type of the Coulomb ground state 
wave function.  Although this simple trial function has been used before, to the best of our knowledge the minimization condition, a third-order algebraic equation,
was solved only approximately (e.g. in the textbook of \cite{GriffithsBook}, see problem 7.14). Here we give its exact solution, and it turns out
that, remarkably, it closely matches our result from fifth-order perturbation theory in the whole range of $\mu$ except the region close to the critical end-point. 

Fourth, in section \ref{sec:mu-crit} we will present a novel, but simple, method to determine the critical screening parameter.
We obtain for it the value

\bear
\frac{\mu_c}{\alpha m} = 1.1906122105(5),
\ear
which is the most precise value to date. We compare with previous results given for the screening parameter.


\section{Review of the literature}
\label{sec:review}
\renewcommand{\theequation}{2.\arabic{equation}}
\setcounter{equation}{0}

A number of general theorems exist on the existence and number of bound states for a given potential. 
In 1951 Pais and Jost \cite{JostPais} showed that for a 3-dimensional spherical potential such that $I=2m\int_0^\infty dr\, r|V(r)|$ is finite, 
bound states must exist for $I>1$.  
One year later, Bargmann \cite{Bargmann} proved that the number of bound states, $n_l$, for a given angular momentum quantum number, $l$, is bounded by

\begin{equation}
 \label{ec:nume}
 (2l+1)n_l<I \, .
\end{equation}
For the Yukawa potential, this relation (in our units) is $(2l+1)n_l<\frac{2m}{\mu}\alpha$. In particular, no bound state
can exist for $\mu > 2m\alpha$. The inequality \eqref{ec:nume} was rederived and further generalized in 1960 by Schwinger \cite{Schwinger}.

As was mentioned in the introduction, no exact results exist to date for the wave functions and energies of the Yukawa potential. 
As to approximative calculations, the most widely used method has been the variational principle. In 1962, 
Harris \cite{Harris} used trial wave functions constructed from the 1s, 2s and 3s solutions of the Coulomb potential 
to obtain very good values for the ground state and the first 45 excited energies of the system. 
In 1990, Garavelli and Oliveira \cite{Garavelli} applied the variational method using the 1s Coulomb solution 
together with a second wave function involving a screening parameter to be determined. 
In 1993, Gomes et al. \cite{Gomes} devised a two-step procedure where optimized few-parameter trial functions are obtained from
an initial linear combination of atomic orbitals (LCAO) with up to 26 basis functions. This allowed them to obtain very precise values
for $E_0(\mu)$ and $\mu_c$, and also to demonstrate the delocalization of the ground state wave function in the bound-unbound transition, i.e.
$\psi_0(r)\to 0$ for $\mu\to \mu_c$. They were also able to determine the critical exponents for $\psi_0^2(0)$ and $E_0$ for this transition. 

Somewhat less popular in this context has been perturbation theory.
The work by Harris already cited \cite{Harris} was also the first to treat the Yukawa potential as a perturbation
of the Coulomb one, although only in first-order perturbation theory. 
G\"on\"ul et al. \cite{Gonul} in 2006 combined the perturbative treatment with an expansion of the
exponential factor $\textrm{e}^{-\mu r}$.
In 1985, Dutt et al. \cite{Dutt} used a scaled Hulth\'en potential instead of the Coulomb one as the unperturbed Hamiltonian.

As to numerical approximations, 
in 1970, Rogers et al. \cite{Rogers} solved the Schr\"odinger equation numerically. 
For the same purpose, in 2005 Yongyao et al. \cite{Yongyao} used Runge-Kutta and Numerov algorithms, as well as Monte Carlo methods.

There have also been less standard approximations to analyze the Yukawa potential. Garavelli and Oliveira \cite{Garavelli} used an iterative process to solve the 
Schr\"odinger equation in momentum space. In 2012, Hamzavi et al. \cite{Hamzavi} used the generalized parametric Nikiforov-Uvarov method for obtaining 
approximate analytical solutions of the Schr\"odinger equation, and showed that this works well for $\mu\lesssim 0.15m\alpha$.

\begin{figure}[t]
\centering
\includegraphics[width=0.50\textwidth]{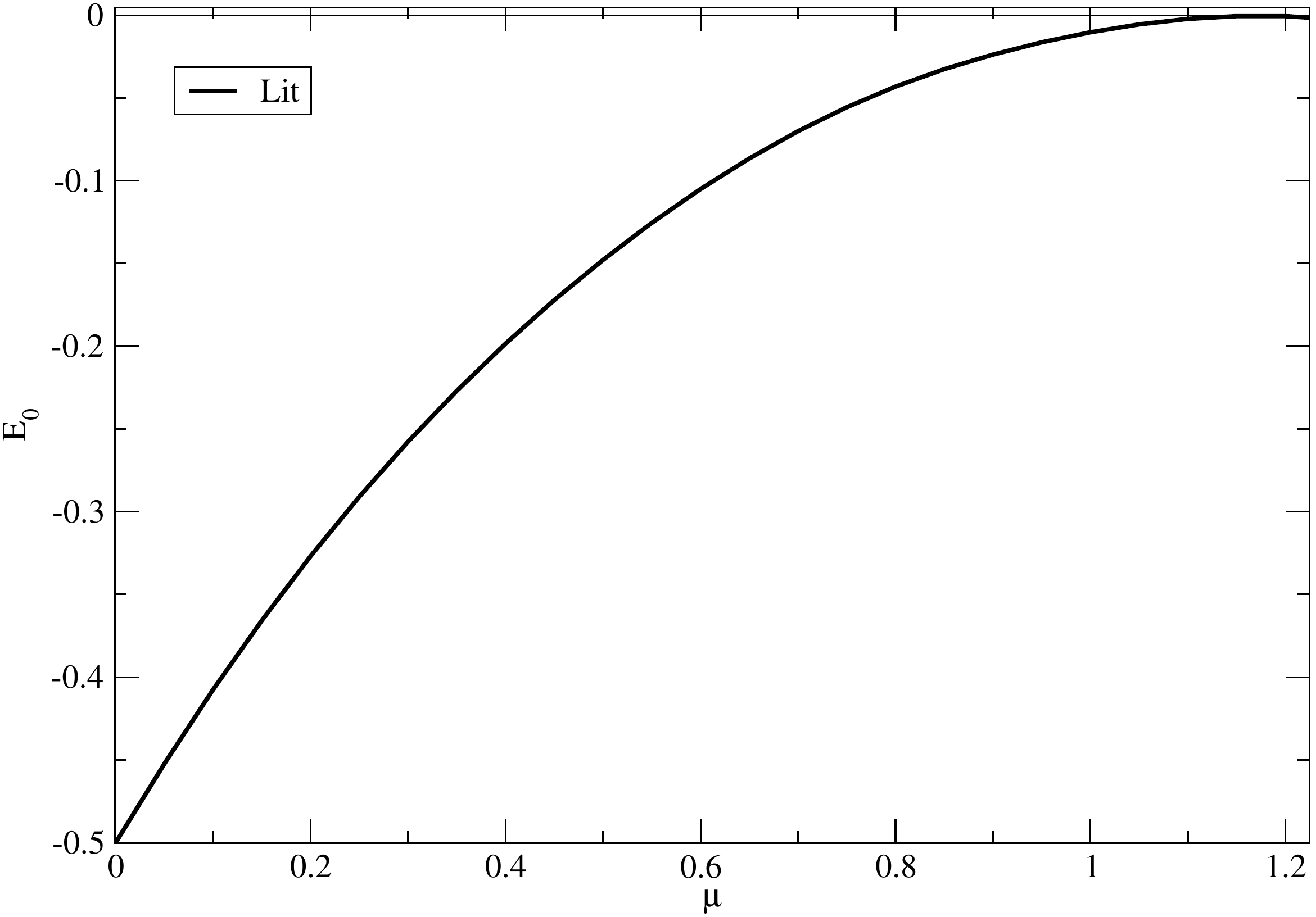}
\caption{A literature average of the Yukawa ground state energy $E_0(\mu)$  (for $m=\alpha =1$).}
\label{fig-YukawaLitAverage}
\end{figure}

In Fig. \ref{fig-YukawaLitAverage} we show a plot of $E_0(\mu)$ obtained by averaging over the results given by various authors, based on TABLE I of \cite{Garavelli}
(we have not included here the results of \cite{Gomes}, since they consider only four different values of $\mu$).

\noindent
The $\mu_c$ values given by a number of authors are listed in Table \ref{table:mu-crit} of section \ref{sec:mu-crit} below.

\section{Perturbative calculation of the ground state energy}
\label{sec:pt}
\renewcommand{\theequation}{3.\arabic{equation}}
\setcounter{equation}{0}

Of the many ways of finding approximate solutions to the Schr\"odinger equation for a system that cannot be solved exactly,  
probably the most widely used is perturbation theory, where one builds on the known exact solutions of some other, usually
simpler system. However, the perturbative expansion becomes quickly cumbersome at higher order, so that most textbooks of quantum
mechanics, e.g. Griffiths \cite{GriffithsBook}, give the explicit formulas only to second order. Exceptionally, Landau and Lifshitz \cite{Landau1Book} 
give them to third order, the fourth order is worked out in an unpublished article \cite{Wheeler}, and Wikipedia \cite{wikiP} has the expressions for the
energy levels to fifth order (for the non-degenerate case). Those expressions, whose correctness we have verified by an independent calculation, 
are included in appendix \ref{app-explicitP} for easy reference.  
Here we wish to apply them to the ground state energy of the Yukawa Hamiltonian, taking advantage of the fact that the Coulomb case is exactly solvable: 
\begin{eqnarray}
H_{\rm Yuk} \equiv-\frac{1}{2m}\nabla^2-\alpha\frac{\textrm{e}^{-\mu r}}{r} = \underbrace{-\frac{1}{2m}\nabla^2-\frac{\alpha}{r} }_{H_0}+
\underbrace{\frac{\alpha}{r}(1-\textrm{e}^{-\mu r})}_{\Delta H}\,.
\label{hamsplit}
\end{eqnarray}
We recall that the eigenvalues of $H_0$ are
\begin{eqnarray}
\label{eq:hyd-energy}
 E_n&=&-\frac{m\alpha^2}{2}\frac{1}{n^2}  \, ,
\end{eqnarray}
with eigenfunctions
\begin{equation}
 \label{eq:Hyd-totsol}
 \psi_{nlm}(r,\theta,\phi)=\sqrt{\left(\frac{2}{na_0}\right)^3\frac{(n-l-1)!}{2n(n+l)!}}\textrm{e}^{-\frac{r}{n a_0}}\left( \frac{2r}{n a_0} \right)^l L_{n-l-1}^{2l+1}
 \left( \frac{2r}{n a_0} \right) Y_l^m(\theta,\phi),
\end{equation}
where $n=1,2,3,\dots$; $l=0,1,2,\cdots,n-1$; $m=-l,\cdots,l$, and $a_0=1/m\alpha$ is the Bohr radius. 
$Y_l^m(\theta,\phi)$ are the spherical harmonics, and $L_n^k$ the associated Laguerre polynomials
(our convention for the latter is given in (\ref{eq:laguere-exp}) in the appendix and corresponds to that of MATHEMATICA).
Since both the unperturbed ground state and the perturbation $\Delta H$ are spherically symmetric, it is easily seen that the eigenstates with non-vanishing angular momentum,
i.e. with $l >0$, will not contribute to any order in the perturbative expansion. This greatly simplifies the expansion, and in particular reduces it to the non-degenerate case,
so that we can use the formulas for non-degenerate perturbation theory as given in appendix~\ref{app-explicitP}, restricting them to the spherically
symmetric eigenfunctions $\psi_{n} \equiv \psi_{n00}$ from the beginning. They involve, apart from the energy differences  $\Delta_{nm}\equiv E_{n}-E_{m}$,
only the matrix elements  $V_{nm}\equiv \langle \psi_{n}|\Delta H|\psi_{m}\rangle$, which we obtain in closed form in \eqref{eq:psinHpsip}.
From the second order correction onwards the expressions involve infinite sums over the principal quantum number $n$, which we were unable to do in closed form.
However, all these sums converge very rapidly (at least as $1/n^3$), so that a cutoff could be used on them; using C++, we were able to sum over the first $19$
terms for each infinite sum. 

\begin{figure}[t!]
\centering
\includegraphics[scale=0.4]{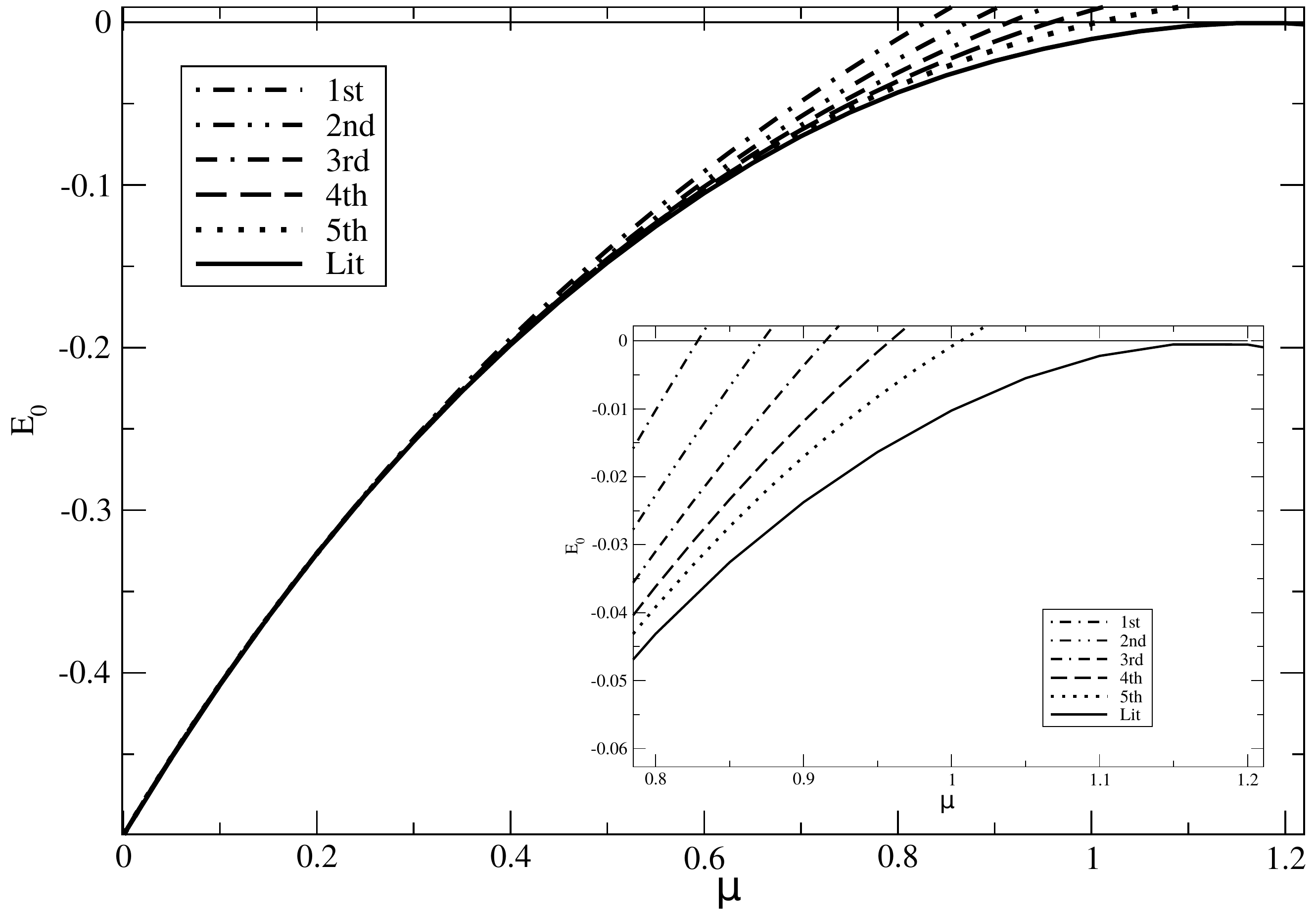}
\caption{The ground state energy $E_0(\mu)$  of the Yukawa Hamiltonian at various orders of perturbation theory (for $m=\alpha =1$), with a cutoff at $n=19$, together
with the literature average curve.}
\label{fig-Hyuk19terms}
\end{figure}

In Fig. \ref{fig-Hyuk19terms} we show a plot of the results of this calculation for the ground-state energy as a function of $\mu$ 
at various orders of perturbation theory (choosing $m=\alpha =1$), together with the literature average curve obtained in the previous section. 
We observe that perturbation theory works well up to $\mu \simeq 0.5$ and breaks down for $\mu \gtrsim 0.8$. As one would expect, 
the addition of the higher-order terms delays the onset of this breakdown, but not very significantly. It is also interesting to note that, in the range where
perturbation theory works, the perturbation series for fixed $\mu$ still shows an apparent convergent behavior to fifth order, even though it is known that 
the perturbation series in quantum mechanics (as well as in quantum field theory) is generically an asymptotic divergent one (see, e.g. \cite{zinnjustin-book}).
It would be interesting to push this calculation to even higher orders to see the onset of asymptoticity.  

\begin{figure}[t!]
\centering
\includegraphics[scale=0.4]{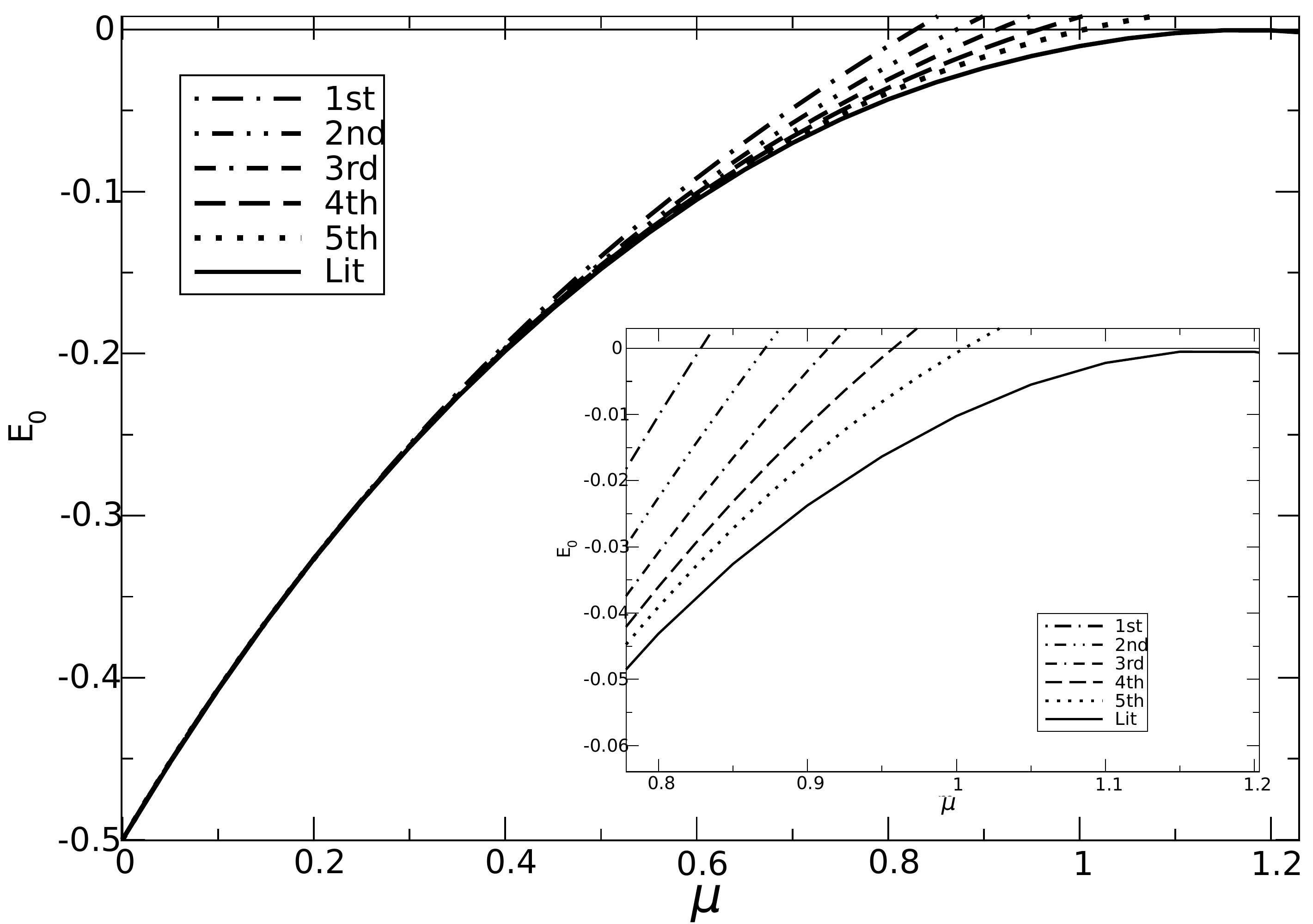}
\caption{The ground state energy $E_0(\mu)$  of the Yukawa Hamiltonian at various orders of perturbation theory (for $m=\alpha =1$), with a cutoff at $n=10$, together
with the literature average curve.}
\label{fig-Hyuk10terms}
\end{figure}

As a check on the cutoff that we have used for the principal number summations, let us also show in Fig. \ref{fig-Hyuk10terms} the corresponding plot obtained by summing 
only over the first
$10$ terms, rather than $19$, for all these sums. The plots are indistinguable in the range of $\mu$ where perturbation theory works. 

\begin{figure}[t!]
\centering
\includegraphics[scale=0.45]{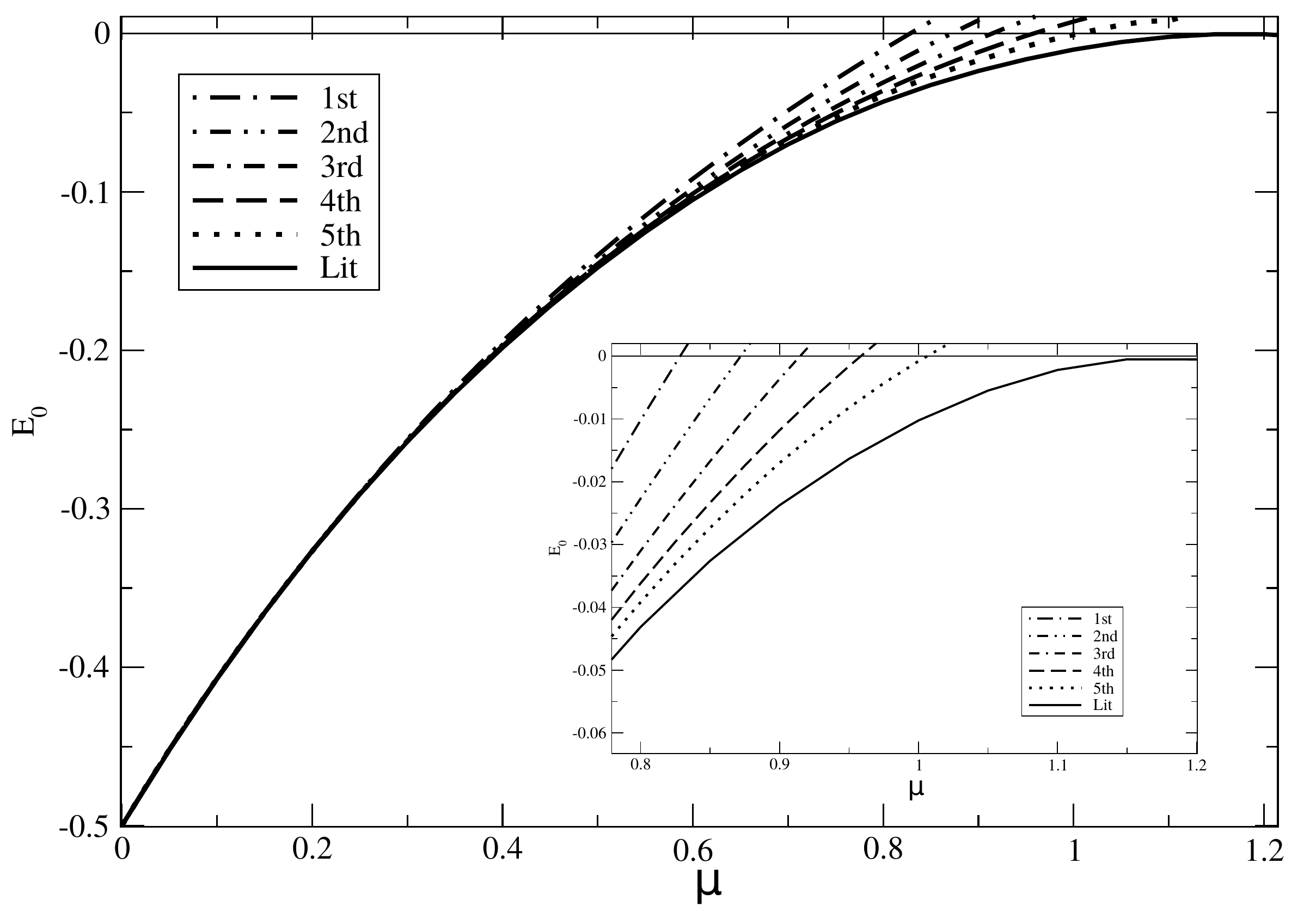}
\caption{The ground state energy $E_0(\mu)$  of the Yukawa Hamiltonian using $H_0'$ (for $m=\alpha =1$), together
with the literature average curve, with a cutoff at $n=19$.}
\label{fig-Hprime19terms}
\end{figure}

The expansion in $\Delta H$ is effectively an expansion in $\mu$, which suggests that better results might be obtained by using the 
same perturbation series with a different break-up of the Yukawa Hamiltonian: instead of \eqref{hamsplit}, let us try moving the
term linear in $\mu$ contained in the Yukawa potential, which corresponds to a constant term in the Hamiltonian, from 
$\Delta H$ to $H_0$:
\begin{eqnarray}
H_{\textrm{Yuk}} =\underbrace{-\frac{1}{2m}\nabla^2-\frac{\alpha}{r}+\mu\alpha
}_{H_0'}+ \underbrace{\frac{\alpha}{r}(1-\mu r - \textrm{e}^{-\mu r})}_{\Delta H'}\, .
\label{hamsplitprime}
\end{eqnarray}
The new $H_0'$ has the same eigenfunctions as $H_0$, and eigenvalues shifted by the constant $\mu\alpha$:
\begin{eqnarray}
\label{eq:hyd-energyprime}
E'_n&=&-\frac{m\alpha^2}{2}\frac{1}{n^2} + \mu \alpha \,.
\end{eqnarray}
Redoing the perturbative calculation with this modification, up to fifth order and with the same cutoff at $n=19$, we have obtained the results for $E_0(\mu)$ shown in Fig.
\ref{fig-Hprime19terms}.

Comparing with Fig. \ref{fig-Hyuk19terms} we see that the new break-up does not make any improvement. 
    An important remark is that increasing the cut-off above $n= 19$ leads to
numerical instability in the fifth order perturbation theory due to delicate cancellations between large numbers (generated by the combinatoric factorials present in the
equations in the appendix). Various ad-hoc modifications to the methods of calculating the summands become necessary to alleviate these problems.

\section{Variational principle}
\label{sec:variational}
\renewcommand{\theequation}{4.\arabic{equation}}
\setcounter{equation}{0}

For the case of the ground state energy, the variational principle provides an approximation method that is more universally applicable than
perturbation theory, and often yields accurate results with relatively little effort. It states that to obtain the ground state energy 
of a system described by the Hamiltonian, $H$, in the case that it is not possible to solve the Schr\"odinger equation exactly, then one can pick any normalized wave 
function $\psi$ whatsoever and,
using the spectral representation of the Hamiltonian, it is easily seen that one always has
\begin{equation}
E_{0}\leq \langle \psi|H|\psi \rangle.
\label{E0ineq}
\end{equation}
That is, unless $\psi$ is the true ground state, the expectation value of $H$ in the state $\psi$ is certain to overestimate the ground state energy \cite{GriffithsBook}.

For the Yukawa Hamiltonian \eqref{ec:yukawa}, the simplest possible choice is a trial wave function $ \psi_{t}(r)$ that 
mimics the ground state wave function of the Coulomb potential, that is the $\psi_{100}$ of equation (\ref{eq:Hyd-totsol}):
\begin{equation}
 \label{eq:Hyd-trial}
 \psi_{t}(r) \equiv \frac{1}{\sqrt{\pi b^3}}\textrm{e}^{-\frac{r}{b}}\,.
\end{equation}
This wave function is normalized, and $b$ is the variational parameter that needs to be adjusted to minimize the expectation value of the Hamiltonian (\ref{eq:Hyd-totsol}).
Since there is only radial dependence, the calculations are simple 
and lead to
\begin{equation}
\label{eq:ener-hyd}
 E_{t} \equiv\langle \psi_{t}(r)|H| \psi_{t}(r) \rangle = \frac{1}{2mb^2}-\frac{4}{a_0 m b(2+b\mu)^2}\,,
\end{equation}
where we have also used the relation $\alpha=1/(m a_0)$.
 
\begin{figure}[t!]
\centering
 \includegraphics[scale=0.4]{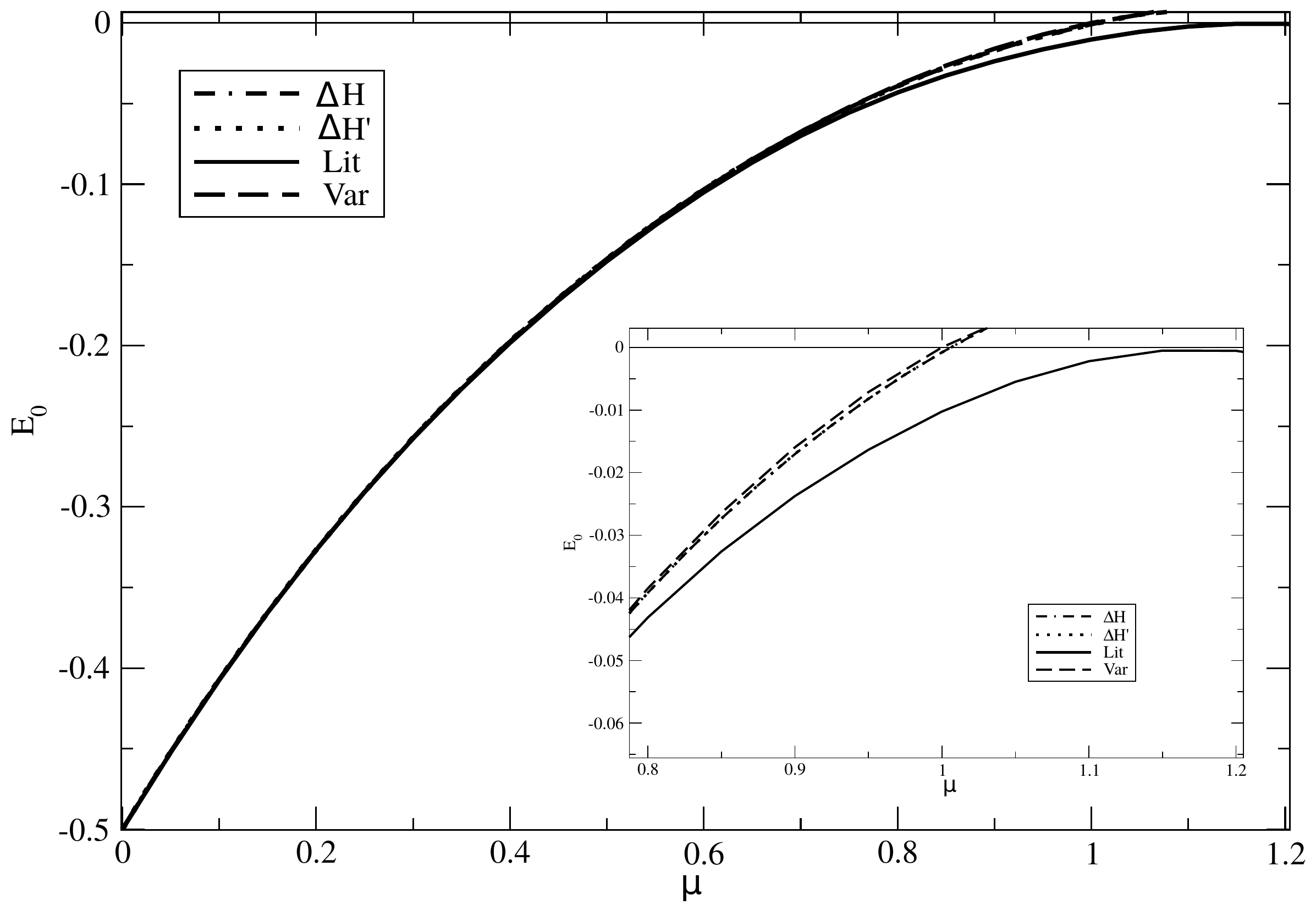}
 \caption{Variational solution to the ground state energy of the Yukawa potential, together with the literature average curve and
 the fifth-order perturbative ones.}
\label{fig-comparevar}
\end{figure}

The minimization of the expression (\ref{eq:ener-hyd}) leads to an algebraic third-order equation. 
Curiously, although the variational method has been applied to the Yukawa potential by a number of authors, 
and the trial function \eqref{eq:Hyd-trial} was used in \cite{GriffithsBook} (in problem 7.14), the exact solution of this
equation seems to be not in the literature. MATHEMATICA gives it as
\setlength\arraycolsep{2pt}{
\begin{eqnarray}
 \label{eq:b0hyd}
 b_0(\mu)&=&\frac{1}{3a_0 \mu^3}\Big[ -6\mu(-2+a_0\mu)+\frac{i 6^{2/3}(i+\sqrt{3})\mu^2(-6+5a_0\mu)}{\left( -36\mu^3+45a_0\mu^4-9a_0\mu^5
 +a_0\mu^4\sqrt{3(-9-20a_0\mu+27a_0\mu^2)}
 \right)^{1/3}}\nonumber\\
 &+& 6^{1/3}(1+i\sqrt{3})\left( -36\mu^3+45a_0\mu^4-9a_0\mu^5+a_0\mu^4\sqrt{3(-9-20a_0\mu+27a_0\mu^2)}\right)^{1/3}\Big] \, .
\end{eqnarray}}
This solution is real, although it is not obviously so. 
The physically relevant one among the three solutions of the third-order equation was determined by assuming that 
$b_0(\mu)$ should go to the Coulomb value $a_0$ for $\mu \to 0$. Thus for $\mu\ll1$ it expands out as $a_0$ plus 
perturbations in powers of $\mu$:

\begin{equation}
 \label{eq:b0hydapp}
 b_0=a_0+\frac{3}{4}a_0^3\mu^2-a_0^4\mu^3+\frac{21}{8}a_0^5\mu^4 - 6a_0^6\mu^5 + \mathcal{O}(\mu^6)\,.
\end{equation}
In Fig. \ref{fig-comparevar} we show the result of using \eqref{eq:b0hyd} in (\ref{eq:ener-hyd}), together with the
literature average curve and the curves from both versions of fifth-order perturbation theory (the curves for $H_0$ and $H_0'$  
are practically indistinguishable at the scale of the figure). 
Note that, for very small $\mu$, the numerical evaluation of the variational curve becomes unstable; a smooth
result can be obtained by replacing the exact $b_0(\mu)$ of \eqref{eq:b0hyd} by its small $\mu$ approximation \eqref{eq:b0hydapp}.
Remarkably, the variational and the fifth-order perturbative curves are in close agreement.

\section{The critical screening parameter $\mu_c$}
\label{sec:mu-crit}
\renewcommand{\theequation}{5.\arabic{equation}}
\setcounter{equation}{0}

As we mentioned in the introduction, an important difference between the Coulomb and Yukawa cases is that, for the latter, bound states cease to exist 
if $\mu$ becomes larger than a critical value $\mu_c$. Such a transition in quantum mechanics holds important information on the dynamics of the system. 
For example, in solid state physics the existence of bound states makes it possible to condense electrons around protons and get an insulating system, 
while in their absence the system is a metal \cite{KittelBook}. The transition between both regimes is called Mott transition. 
The value of $\mu_c=(\cdot)m\alpha$ is not known exactly. In  Table \ref{table:mu-crit} we give some values for $(\cdot)$ found in the literature, together with the ones 
following from our approximate calculations above, determined by the intersection of the $E_0(\mu)$ curve with the $E_0 = 0$ axis.

\begin{table}[t!]
\centering
\caption{Comparison of $(\cdot)$ values for different works and our results. The letters in parentheses indicate the method used;
V for variational, VC variational 
taking as trial wave function the ground state wave function of the Coulomb potential, P perturbation theory and N numerical, A analytical method by GO \cite{Garavelli},
LCAO (linear combination of atomic orbits) by Gomes et. al. \cite{Gomes}.}

\bigskip

\begin{tabular}{|c||c||c||c||c|}
\hline
 Harris \cite{Harris} (V) & Harris \cite{Harris} (P) & GO \cite{Garavelli} (A) & GO \cite{Garavelli} (VC) & GO  \cite{Garavelli} (V) \\
 \hline 
 1.15 & 0.828 & 1.189621 & 1.0 & 1.190213  \\
\hline	
\hline
GCM \cite{Gomes} (V) & GCM \cite{Gomes} (LCAO) &Harris \cite{Harris} (VC) & RGH \cite{Rogers} (N) & YXK \cite{Yongyao} (N)  \\ 
\hline
 1.19061074 & 1.19061227 & 1.0 & 1.190607 & 1.1906\\
\hline
\hline
 HL \cite{Hulthen} & Us (VC) & Us (P with $H_0'$) & Us (P with $H_0$) & \\
\hline
 1.1906 & 1.0 & 1.006 & 1.006 & \\
\hline
\end{tabular}
\label{table:mu-crit}
\end{table}

Further, we will now present a method for the calculation of $\mu_c$ that is quite simple, but which we have not been able to find in the literature. 
For $\mu=\mu_c$ we have the lowest eigenvalue $E_0 =0$ with corresponding radial Schr\"odinger equation 
\bear
-\frac{1}{2m}\Bigl(\frac{d^2}{dr^2} + \frac{2}{r} \frac{d}{dr} \Bigr) \psi_0(r) -\frac{\alpha}{r}{\rm e}^{-\mu r} \psi_0(r) = 0\,.
\label{schroed}
\ear
In terms of $g(r) \equiv r\psi_0(r)$, this can be written as
\bear
g''(r) + \alpha m \frac{2}{r}{\rm e}^{-\mu r}  g(r) = 0\,.
\label{schroed2}
\ear
According to \eqref{mucrit} the critical value $\mu_c$ is proportional to $\alpha m$, so that we can as well set $\alpha m =1$. 
Equation \eqref{schroed2} describes, for $\mu$ just below the critical value $\mu_c$, a bound state solution, and for $\mu$ just above
$\mu_c$ a scattering solution (here we neglect an infinitesimal contribution to the right hand side of (\ref{schroed2}), $-2mE_{0} g(r)$, that is present whilst $\mu \ne \mu_{c}$).
In either case we must have $g(0)=0$, for regularity of the wave function at the origin, and in the bound state case
also $g(\infty) = 0$, since the wave function must decrease (much) faster than $1/r$ for large $r$ to be square-integrable. 
This suggests \cite{Hulthen} that one could distinguish between $\mu <\mu_c$ and $\mu >\mu_c$ by checking numerically whether \eqref{schroed2}
can be solved with the boundary conditions 

\bear
g(0) = g(\infty) = 0\,.
\label{bc1}
\ear
However, these boundary conditions are not the most convenient ones for
numerical purposes. It is thus important to observe that we can as well replace them by the conditions 

\bear
g(0) = 0,\, g'(0) = g_1\,,
\label{bc2}
\ear
where $g_1$ is an arbitrary non-zero number, and check whether the numerical solution of  \eqref{schroed2} leads to $g(\infty) =0$ or not. 
These procedures are equivalent, since the boundary conditions \eqref{bc1} are homogeneous, so that a solution of \eqref{schroed2}
fulfilling them can be rescaled to make $g'(0)$ take any given non-zero value $g_{1}$ (non-zero, since $g'(0) = g(0)=0$ leads to the trivial solution
$g(0) \equiv 0$). Moreover, it is easy to check the behaviour of $g(r)$ at large $r$, since \eqref{schroed2} implies that $g'(r)$ rapidly converges to
a constant for large $r$. Thus, the critical $\mu_c$ can be found by starting with a $\mu$ somewhat below it, and then hiking it up little by little,
each time solving \eqref{schroed2} numerically with some arbitrary $g_1>0$, and checking whether the slope of $g(r)$ still becomes negative for large $r$,
as it must for $g(r)$ to represent a bound state solution (in the bound state case, the slope of $g(r)$ ultimately must go to zero for very large $r$; in an exact calculation,
this would be ensured by the tiny positive contribution to the right hand side of (\ref{schroed2}) that is present whilst $\mu < \mu_{c}$).

\begin{figure}[t!]
\centering
 \includegraphics[scale=0.5]{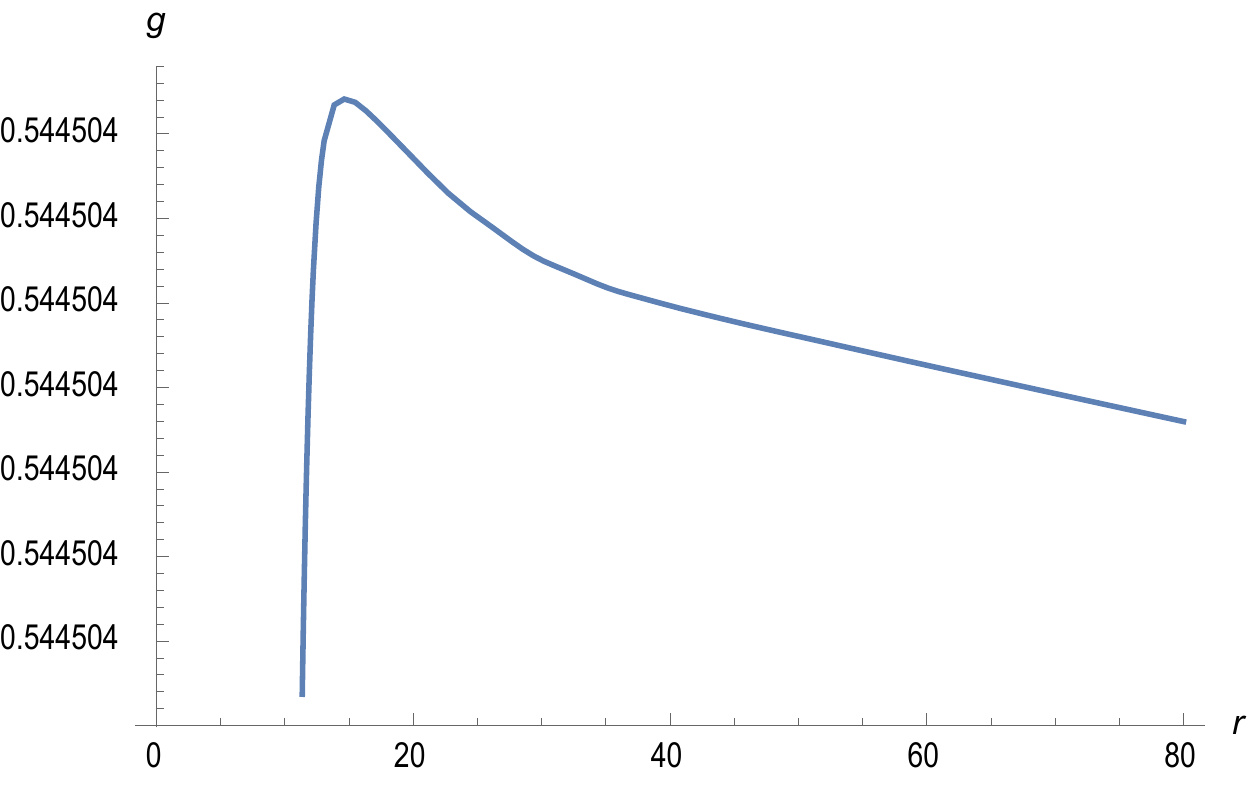}
 \caption{MATHEMATICA plot of $g(r)$ for $\mu=1.190612210$, just below its critical value.}
\label{fig-yukawabelow}
\end{figure}

\begin{figure}[t!]
\centering
 \includegraphics[scale=0.5]{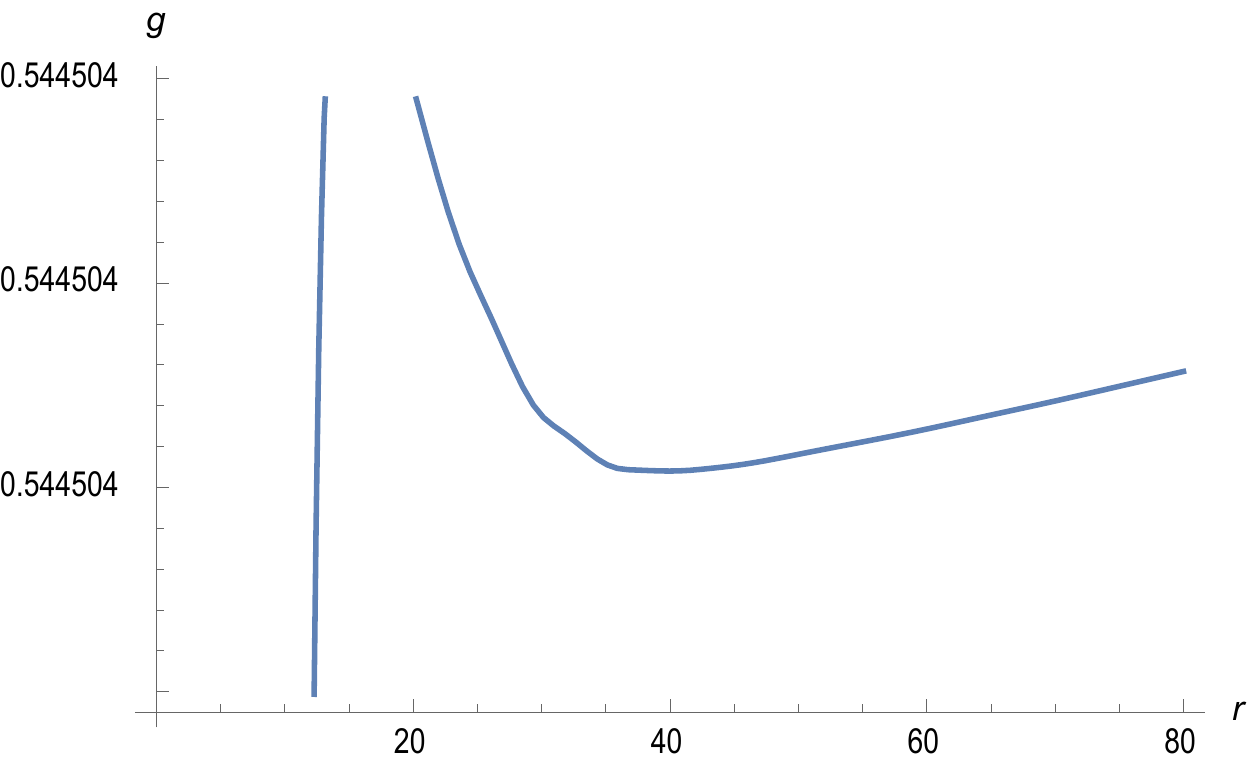}
 \caption{MATHEMATICA plot of $g(r)$ for $\mu=1.190612211$, just above its critical value.}
\label{fig-yukawaabove}
\end{figure}

 In Figs. \ref{fig-yukawabelow} and \ref{fig-yukawaabove} we show plots obtained by
setting $g_1=1$ and using the NDSolve command of MATHEMATICA10. Since $r=0$ cannot be used in the numerical evaluation, here we replaced
the exact conditions \eqref{bc2} by the approximate ones
\bear
g(r_0) = g(0) + r_0g'(0) = r_0 g_1,\,\, g'(r_0) = g'(0) = g_1\,, 
\label{bc2approx}
\ear
with $g_1=1$ and the small radial cutoff $r_0 = 10^{-10}$.

From these plots we conclude that $\mu_c/\alpha m$ lies between $1.190612210$ and $1.190612211$. 
This is compatible with the LCAO result of \cite{Gomes}, $\mu = 1.19061227(4)$ (except for their last digit) and \cite{Yongyao}, but not with \cite{Rogers}
and \cite{Garavelli}. We have also checked
that this result is stable under a further reduction of the radial cutoff $r_0$. 

\section{Conclusions}
\label{sec:conc}
\renewcommand{\theequation}{6.\arabic{equation}}
\setcounter{equation}{0}

In this paper, we have summarized the existing literature on the ground state energy $E_0$ and the critical screening parameter $\mu$ for
the Yukawa potential, and we have performed three calculations that, although straightforward, to the best of our knowledge
have not been done before: (i) a fifth-order perturbative calculation of $E_0$ using the exact solutions of the Coulomb potential together
with a cutoff on the principal number summations (ii) a variational calculation of $E_0$ using a simple Coulomb-like radial wave function 
and the exact solution of the corresponding minimization condition, a third-order equation (iii) a high-precision determination of $\mu_c$ by numerical integration
of the radial Schr\"odinger equation with appropriate boundary conditions. 
Our main findings are the close agreement between the fifth-order perturbative result with the variational one, 
and a calculation
of $\mu_c$ to ten significant numbers. This is one digit more than was obtained in the LCAO  calculation of \cite{Gomes}, previously the most precise determination of
this parameter available in the literature, with an incomparably greater effort. Our calculation essentially confirms their value, 
and shows the inaccuracy of some other values found in the literature. Moreover, since our new approach essentially amounts to a form of interval bisection, its rate of convergence, being linear, is an improvement on the computational complexity of many other approaches, making it easier -- and less time consuming -- to increase the accuracy of our determination of $\mu_{c}$ to greater numbers of decimal places. The result presented here should also become useful as a benchmark for future approximative calculations.
Moreover, the method proposed here for the calculation of $\mu_c$ may find generalizations to other spherically symmetric potentials. 

\appendix   
\section{Formulas needed for the fifth-order perturbation expansion}
\label{app-explicitP}
\renewcommand{\theequation}{A.\arabic{equation}}
\setcounter{equation}{0}

For easy reference, let us list here the expressions given in Wikipedia \cite{wikiP} for the energy level corrections up to fifth order
in (non-degenerate) perturbation theory:

\setlength\arraycolsep{2pt}{
\begin{eqnarray}
\label{eq:energy1stc}
 \Delta E_n^1 &=& V_{nn}\\
 \label{eq:energy2ndc}
 \Delta E_n^2 &=& \sum_{m\neq n}\frac{V_{nm}^2}{\Delta_{nm}}\\
 \label{eq:energy3rdc}
 \Delta E_n^3 &=& \sum_{m\neq n}\sum_{r\neq n}\frac{V_{nr}V_{rm}V_{mn}}{\Delta_{nm}\Delta_{nr}}-V_{nn}\sum_{m\neq n}\frac{V_{nm}^2}{\Delta_{nm}^2}\\
 \label{eq:energy4thc}
 \Delta E_n^4 &=& \sum_{l\neq n}\sum_{m\neq n}\sum_{r\neq n}\frac{V_{nr}V_{rm}V_{ml}V_{ln}}{\Delta_{nr}\Delta_{nm}\Delta_{nl}} - V_{nn}\sum_{l\neq n}\sum_{m\neq n}
 \frac{V_{nm}V_{ml}V_{ln}}{\Delta_{nm}^2\Delta_{nl}}-V_{nn}\sum_{l\neq n}\sum_{r\neq n}\frac{V_{nr}V_{rl}V_{ln}}{\Delta_{nr}\Delta_{nl}^2}\nonumber\\
 && + V_{nn}^2\sum_{l\neq n}\frac{V_{nl}^2}{\Delta_{nl}^3}-\sum_{l\neq n}\sum_{m\neq n}\frac{V_{nm}^2 V_{nl}^2}{\Delta_{nm}\Delta_{nl}^2}
 \end{eqnarray}}
 \setlength\arraycolsep{2pt}{
\begin{eqnarray}
 \label{eq:energy5thc}
 \Delta E_n^5 &=& \sum_{s\neq n}\sum_{l\neq n}\sum_{m\neq n}\sum_{r\neq n}\frac{V_{nr}V_{rm}V_{ml}V_{ls}V_{sn}}{\Delta_{ns}\Delta_{nm}\Delta_{nl}\Delta_{nr}} - 
 \sum_{s\neq n}\sum_{l\neq n}\sum_{m\neq n}\frac{V_{nm}^2 V_{nl}V_{ls}V_{sn}}{\Delta_{nm}\Delta_{nl}^2\Delta_{ns}} \nonumber\\
 &&- \sum_{s\neq m}\sum_{l\neq n}\sum_{r\neq n}\frac{V_{nl}^2 V_{nr}V_{rs}V_{sn}}{\Delta_{nr}\Delta_{ns}^2 \Delta_{nl}}
  -\sum_{s\neq n}\sum_{m\neq n}\sum_{r\neq n}\frac{V_{ns}^2 V_{nr}V_{rm}V_{mn}}{\Delta_{ns}^2\Delta_{nm}\Delta_{nr}}\nonumber\\
 &&- V_{nn}\sum_{s\neq n}\sum_{l\neq n}
 \sum_{m\neq n}\frac{V_{nm}V_{ml}V_{ls}V_{sn}}{\Delta_{nm}^2 \Delta_{nl}\Delta_{ns}} - V_{nn}\sum_{s\neq n}\sum_{l\neq n}\sum_{r\neq n}
 \frac{V_{nr}V_{rl}V_{ls}V_{sn}}{\Delta_{nr} \Delta_{nl}^2 \Delta_{ns}}\nonumber\\
 && - V_{nn}\sum_{s\neq n}\sum_{m\neq n}\sum_{r\neq n}\frac{V_{nr}V_{rm}V_{ms}V_{sn}}{\Delta_{nr} \Delta_{nm}\Delta_{ns}^2} + V_{nn}\sum_{s\neq n}\sum_{m\neq n}
 \frac{V_{nm}^2 V_{ns}^2}{\Delta_{nm}^3 \Delta_{ns}} + V_{nn}\sum_{s\neq n}\sum_{m\neq n}\frac{V_{nm}^2 V_{ns}^2}{\Delta_{nm} \Delta_{ns}^3}\nonumber\\
 && + V_{nn}\sum_{s\neq n}\sum_{m\neq n}\frac{V_{nm}^2 V_{ns}^2}{\Delta_{nm}^2 \Delta_{ns}^2}
 + V_{nn}^2\sum_{s\neq n}\sum_{l\neq n}\frac{V_{nl} V_{ls} V_{sn}}{\Delta_{nl}^3 \Delta_{ns}}
 + V_{nn}^2\sum_{s\neq n}\sum_{m\neq n}\frac{V_{nm}V_{ms}V_{sn}}{\Delta_{nm}^2 \Delta_{ns}^2} \nonumber\\
&& + V_{nn}^2\sum_{s\neq n}\sum_{r\neq n}\frac{V_{nr}V_{rs}V_{sn}}{\Delta_{nr} \Delta_{ns}^3}-V_{nn}^3\sum_{s\neq n}\frac{V_{ns}^2}{\Delta_{ns}^4}
\end{eqnarray}}
Here we have used the short-hand notation $V_{nm}\equiv \langle \psi_{n}|\Delta H|\psi_{m}\rangle$ and $\Delta_{nm}\equiv E_{n}-E_{m}$.

\noindent
Using the formula for the associated Laguerre polynomials
\begin{equation}
 \label{eq:laguere-exp}
 L_n^{b}(x)=\sum_{i=0}^{n}{{n+b}\choose{n-i}}\frac{(-x)^i}{i!}\,,
\end{equation}
we can write

\begin{equation}
 \label{eq:eq:psi_n00exp}
 \psi_{n00}(r)=\frac{1}{n^2\sqrt{n\pi a_0^3}}\textrm{e}^{-\frac{r}{n a_0}}\sum_{i=0}^{n-1} {{n}\choose{n-1-i}}\frac{1}{i!}
 \left(- \frac{2r}{n a_0} \right)^i\,.
\end{equation}
Considering first the case of $\Delta H' = \frac{\alpha}{r}(1-\mu r - \textrm{e}^{-\mu r})$, here we find

\setlength\arraycolsep{2pt}{
\begin{eqnarray}
 \label{eq:psinHpsip}
\langle\psi_{n00}|\Delta H'|\psi_{p00}\rangle &=& \!\frac{4\pi\alpha}{n^2p^2\sqrt{np}\pi a_0^3}\sum_{i=0}^{n-1}\sum_{j=0}^{p-1}{{n-1+1}\choose{n-1-i}}{{p-1+1}\choose{p-1-j}}
 \frac{1}{i!}\frac{1}{j!}\nonumber\\ 
 &&\int_0^{\infty}dr r\textrm{e}^{-\frac{r}{na_0}}\textrm{e}^{-\frac{r}{pa_0}}(1-\textrm{e}^{-\mu r}-\mu r)
 \left(-\frac{2r}{n a_0} \right)^i\left(- \frac{2r}{pa_0}\right)^j\nonumber\\
 &&\hspace{-2cm} = \frac{4\alpha}{n^2p^2\sqrt{np}a_0^3}\sum_{i=0}^{n-1}\sum_{j=0}^{p-1}{{n}\choose{n-1-i}}{{p}\choose{p-1-j}}
 \frac{1}{i!}\frac{1}{j!}\left(-\frac{2}{n a_0} \right)^i\left(- \frac{2}{pa_0}\right)^j\nonumber\\
 &&\int_0^{\infty}dr r^{i+j+1}\left[\textrm{e}^{-r\left( \frac{1}{na_0}+
 \frac{1}{pa_0} \right)}- \textrm{e}^{-r\left( \frac{1}{na_0}+\frac{1}{pa_0}+\mu \right)}-\mu r \textrm{e}^{-r\left( \frac{1}{na_0}+
 \frac{1}{pa_0} \right)}\right]\nonumber\\
  &&\hspace{-2cm} = \frac{4\alpha}{n^2p^2\sqrt{np}a_0^3}\!\sum_{i=0}^{n-1}\!\sum_{j=0}^{p-1}{{n}\choose{n-1-i}}\!{{p}\choose{p-1-j}}\!
 \left(-\frac{2}{n a_0} \right)^i\!\left(- \frac{2}{pa_0}\right)^j\frac{(i+j+1)!}{i!j!}\nonumber\\
 &&\hspace{-1.5cm}\left[\frac{1}{\left( \frac{1}{na_0}+\frac{1}{pa_0} \right)^{i+j+2}}-\frac{1}
{ \left( \frac{1}{na_0}+ \frac{1}{pa_0} +\mu\right)^{i+j+2}}-\mu(i+j+2)\frac{1}{\left( \frac{1}{na_0}+\frac{1}{pa_0} \right)^{i+j+3}}\right]\nonumber\\
&&\hspace{-2cm} = \frac{4m \alpha^2}{\sqrt{np}}\sum_{i=0}^{n-1}\sum_{j=0}^{p-1}{{n}\choose{n-1-i}}{{p}\choose{p-1-j}}\frac{(i+j+1)!}{i! j!}
\Bigg[\frac{1}{(p+n)^2}\left(\frac{-2p}{p+n}\right)^i\left(\frac{-2n}{p+n}\right)^{j}\nonumber\\
&& \hspace{-1.5cm} 
-\frac{m^2\alpha^2}{[(p+n)m\alpha+pn\mu]^2}\left(\frac{-2pm\alpha}{(p+n)m\alpha+pn\mu}\right)^i
\left(\frac{-2nm\alpha}{(p+n)m\alpha+pn\mu}\right)^{j}\nonumber\\
&&\hspace{-1.5cm}-\frac{\mu }{m\alpha}\frac{np}{(p+n)^3}(i+j+2)
\left(\frac{-2p}{p+n}\right)^{i}\left(\frac{-2n}{p+n}\right)^{j}\Bigg]\, ,
\end{eqnarray}}
\noindent
where in particular cases the sums over $i$ and $j$ can be calculated in closed form:

\setlength\arraycolsep{2pt}{
\begin{eqnarray}
\label{eq:psi1Hpsi1}
\langle \psi_{100}|\Delta H'|\psi_{100} \rangle &=& -\frac{\mu^{2}\alpha(3m\alpha+\mu)}{\left(2m\alpha+\mu\right)^2}\,,\\
\label{eq:psinHpsi1}
 \langle\psi_{n00}|\Delta H'|\psi_{100} \rangle &=& 4m\alpha^2\frac{\sqrt{n}}{(n+1)^2}\nonumber\\
 &&\hspace{-6.0cm}\times\left[
 \left(\frac{n-1}{n+1} \right)^{n-1}-\left(\frac{(n+1)m\alpha}{(n+1)m\alpha +n\mu}\right)^2\left(\frac{(n-1)m\alpha+n\mu}{(n+1)m\alpha+n\mu} \right)^{n-1}+
 n(n+1)^{2}\frac{\mu}{8m\alpha}s(n)\right]\,,
\end{eqnarray}}
where
$$s(n)\equiv\begin{cases}
	-2, & n=1 \, , \\
	0, & n>1\, .
\end{cases}$$ 

With these expressions in hand, the corrections to any order to the ground state energy could be calculated in principle. 
The remaining task is to deal with the infinite sums.

To get the respective expressions for $\Delta H =  \frac{\alpha}{r}(1- \textrm{e}^{-\mu r})$, we just have to ignore the last term in the square brackets in
(\ref{eq:psinHpsip}).
The particular cases \eqref{eq:psi1Hpsi1}, \eqref{eq:psinHpsi1} become

\setlength\arraycolsep{2pt}{
\begin{eqnarray}
\label{eq:psi1Hpsi12}
\langle \psi_{100}|\Delta H|\psi_{100} \rangle &=& m\mu\alpha^2\left[\frac{4m\alpha+\mu}{\left(2m\alpha+\mu\right)^2}\right],\\
 \langle\psi_{n00}|\Delta H|\psi_{100} \rangle &=& 4m\alpha^2\frac{\sqrt{n}}{(n+1)^2}\nonumber\\
 &&\hspace{-2.0cm}\times\left[
 \left(\frac{n-1}{n+1} \right)^{n-1}-\left(\frac{(n+1)m\alpha}{(n+1)m\alpha +n\mu}\right)^2\left(\frac{(n-1)m\alpha+n\mu}{(n+1)m\alpha+n\mu} \right)^{n-1}\right]\,
 .\nonumber\\
 \label{eq:psinHpsi12}
\end{eqnarray}}

%
%
%

\end{document}